\documentclass[sigconf]{acmart}
\copyrightyear{2018}
\acmYear{2018}
\setcopyright{iw3c2w3}
\acmConference[WWW 2018]{The 2018 Web Conference}{April 23--27, 2018}{Lyon, France}
\acmBooktitle{WWW 2018: The 2018 Web Conference, April 23--27, 2018, Lyon, France}
\acmPrice{}
\acmDOI{10.1145/3178876.3186129}
\acmISBN{978-1-4503-5639-8/18/04}

\usepackage{graphicx, color}
\usepackage{subfig}
\usepackage{tikz}
\usetikzlibrary{arrows}

\newcommand{\xhdr}[1]{\paragraph*{\bf {#1}.}}

\def\meanout{\mu_{\text{out}}}
\def\meanoutdiff{\delta_{\text{out}}}
\def\gmx{G \backslash x}
\def\flr{f_{\to}}
\def\frl{f_{\gets}}
\newcommand{\rw}[1]{\hat{#1}}
\def\subc{\texttt{r/hillaryclinton}}
\def\subt{\texttt{r/The\_Donald}}
\def\subp{\texttt{r/politics}}
\newcommand{\pst}[1]{s_T(#1)}
\newcommand{\psr}[1]{s_R(#1)}
\def\tran{^\top}
\def\pxct{P_T(x|C)} % Twitter
\def\pxtt{P_T(x|T)} % Twitter
\def\pxcr{P_R(x|C)} % Reddit
\def\pxtr{P_R(x|T)} % Reddit
\def\setc{U_C}
\def\sett{U_T}
\DeclareMathOperator{\vol}{vol}
\newcommand{\E}[1]{\mathbb{E} \left[ #1 \right]}

\title{Mapping the Invocation Structure of Online Political Interaction}
\author{Manish Raghavan}
\affiliation{Cornell University}
\author{Ashton Anderson}
\affiliation{University of Toronto}
\author{Jon Kleinberg}
\affiliation{Cornell University}
\begin{document}
\begin{abstract}
  The surge in political information, discourse, and interaction has
been one of the most important developments in social media over the
past several years. There is rich structure in the 
interaction among different viewpoints on the ideological spectrum.
However, we still have only a limited analytical vocabulary for
expressing the ways in which these viewpoints interact.

In this paper, we develop network-based methods that operate on the
ways in which users share content; we construct \emph{invocation
graphs} on Web domains showing the extent to which pages from one
domain are invoked by users to reply to posts containing pages from
other domains. When we locate the domains on a political spectrum
induced from the data, we obtain an embedded graph showing how these
interaction links span different distances on the spectrum. The
structure of this embedded network, and its evolution over time, helps
us derive macro-level insights about how political interaction
unfolded through 2016, leading up to the US Presidential election. In
particular, we find that the domains invoked in replies spanned
increasing distances on the spectrum over the months approaching the
election, and that there was clear asymmetry between the left-to-right
and right-to-left patterns of linkage.

\end{abstract}
\maketitle

\section{Introduction}
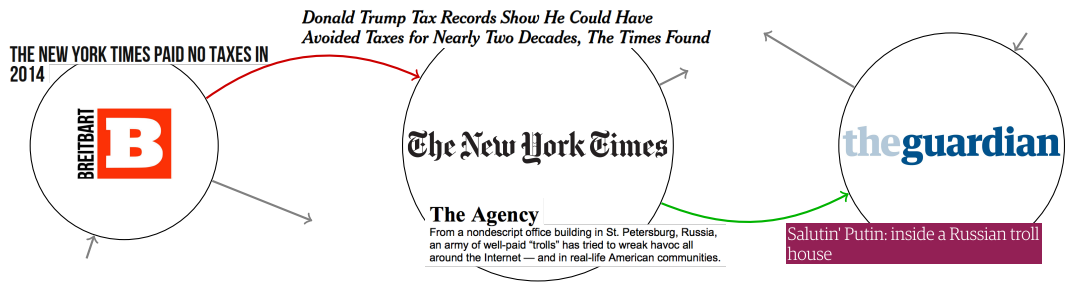
\begin{figure*}[ht]
  \centering
  \begin{tikzpicture}
    \node[circle, draw, minimum size=2.5cm] (bb) at (-1.5, 0) {};
    \node (bbl) at (-1.5, 0) {\includegraphics[width=1.3cm]{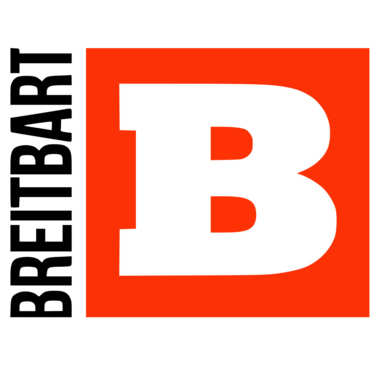}};
    \node[inner sep=0pt] (bba) at (-1.3,1.1)
    {\includegraphics[width=3.5cm]{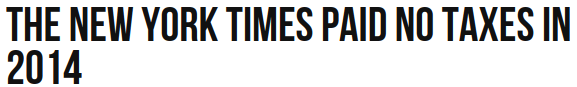}};
    \node[circle, draw,minimum size=3.6cm] (nyt) at (4, 0) {};
    \node (nytl) at (4, 0) {\includegraphics[width=3.6cm]{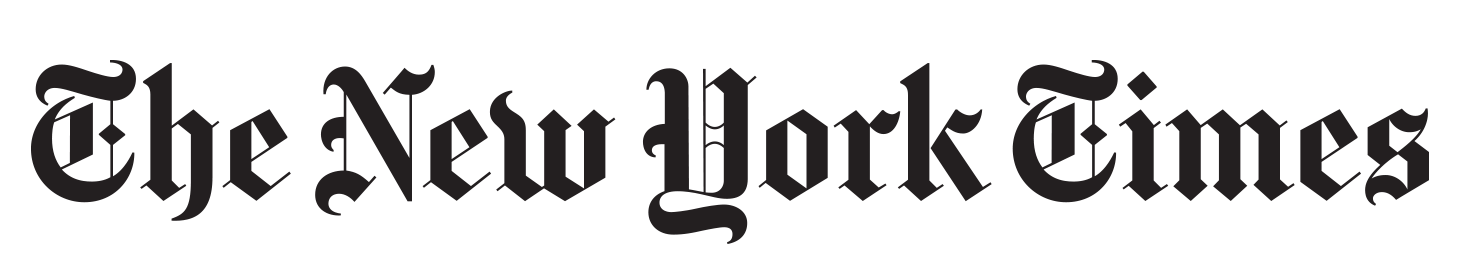}};
    \node (nyta) at (3.6,1.6)
    {\includegraphics[width=5.5cm]{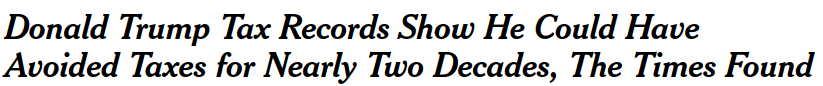}};
    \node (nytt) at (4.5,-1.2)
    {\includegraphics[width=4cm]{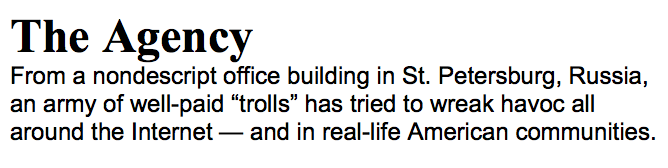}};
    \node (tgl) at (9.5, 0) {\includegraphics[width=3cm]{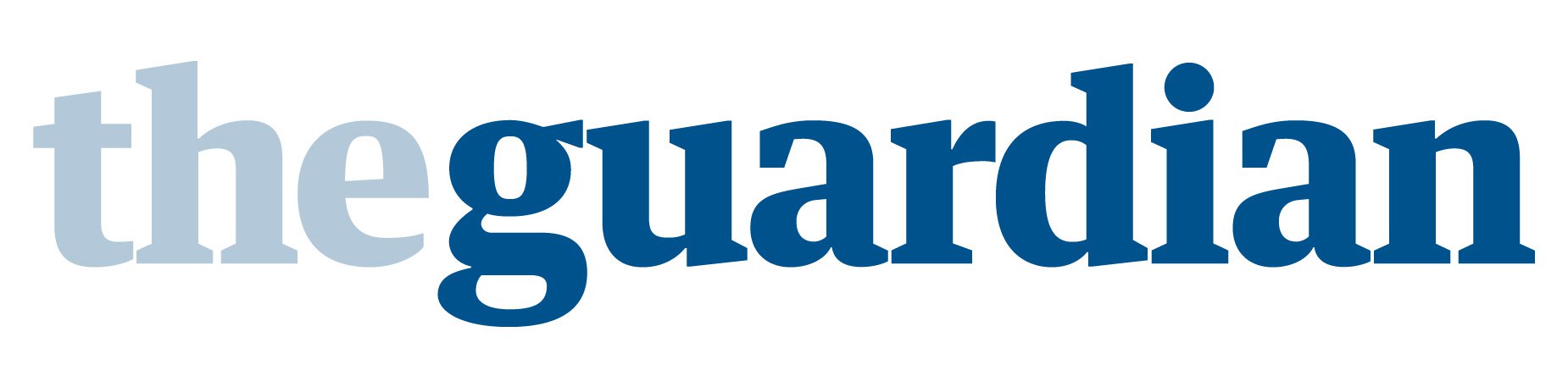}};
    \node[circle, draw,minimum size=3cm] (tg) at (9.5, 0) {};
    \node (tgt) at (9,-1.3)
    {\includegraphics[width=3.4cm]{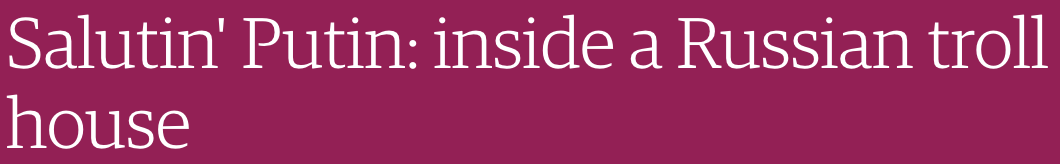}};
    \path[->,black!20!red,thick]
    (bb) edge[bend left=30] node {} (nyt);
    \path[->,black!30!green,thick]
    (nyt) edge[bend right=25] node {} (tg);
    \draw[->, gray, thick] (-2, -1.5) -- (bb);
    \draw[->, gray, thick] (bb) -- (1, -1);
    \draw[->, gray, thick] (tg) -- (7, 1.5);
    \draw[->, gray, thick] (nyt) -- (6, 1);
    \draw[->, gray, thick] (10.5,1.5) -- (tg);
  \end{tikzpicture}
  \caption{A section from an example invocation graph containing instances
  from our data.  In each case, an article from domain $B$ in response
  to an article from domain $A$ contributes to the link from $B$ to $A$.
  The link from Breitbart to The
  New York Times comes from articles like the pair shown here, a Breitbart
  article on the Times' taxes in response to a Times article on Donald Trump's
  taxes. Other links demonstrate an interaction where one article supports
  another: for example, a New York Times article
  shared in response to a Guardian article on Russian information operations
  contributes to the link from The Times to The Guardian.}
  \label{fig:reply_graph}
\end{figure*}

Political interaction has long constituted a key use of 
social media, and there is a correspondingly rich history 
of research into its structure --- one that extends the much longer
history of scholarship on the role of media in the political process 
\cite{bennett-news-illusion,gentzkow-media-slant,kovach-warp-speed,lazarsfeld-peoples-choice}.

 A crucial issue in this line of work is the extent to which 
political interaction on social media takes place primarily 
among users who are ideologically similar, or whether it reaches
across the political spectrum.
Early analysis of political blogging indicated a clustered structure,
with a high density of
linkage among ideologically similar blogs and a lower density of linkage 
between blogs with strongly differing views \cite{adamic-blogosphere}.
Subsequent work, looking at platforms that arose further into
the evolution of social media, suggested that a more complex
structure was developing, in which homophily in views remained
a powerful force, but where the platforms were providing users with some
level of cross-cutting exposure
\cite{bakshy-socnet-info-diffuse,bakshy-ideologically-diverse}.
These questions are important, as they ask whether online political
interaction consists of opposing sides who engage with each other, or 
well-separated clusters who are isolated in ``echo chambers'' or
``filter bubbles'' 
\cite{flaxman-spectrum,pariser-filter-bubble,sunstein-republic-com}.

The answers to such questions 
depend intrinsically on which types of interactions are being considered.
Existing work in the online domain
has implicitly focused on two standard forms of interaction:
page-to-page interaction, expressed by hyperlinks among documents
\cite{adamic-blogosphere};
and user-to-user interaction, expressed by communication among people on
social platforms 
\cite{bakshy-socnet-info-diffuse,bakshy-ideologically-diverse,conover-polarization-twitter}.
Each of these induces a network on a set of entities --- sources and
users respectively --- which can then be analyzed relative to an
underlying political spectrum.

\xhdr{Networks of sources invoked by users}
Here we consider a different type of political interaction network, 
defined as follows.
When a user $u$ shares a page $A$, and a user $v$ replies by 
sharing a page $B$, there is not simply an interaction between
users $u$ and $v$; an interaction is also induced between pages $A$ and $B$.
As reshares develop into a widespread style of social media content production 
\cite{cheng-cascade-prediction,dow-facebook-cascades,goel-structural-virality,kumar-conversations}, 
the ability of users
to deploy page references as proxies in their discussion becomes
an activity requiring very low effort, and 
we find through a large-scale analysis of Twitter data leading up to
the 2016 U.S. Presidential election that such $A$-$B$ interactions
are widespread: users regularly invoke links in this back-and-forth
fashion when they interact with each other.

These {\em invoked interactions} between pages $A$ and $B$ 
are fundamentally different from both 
user-to-user and page-to-page interaction networks.
Unlike free-form user-to-user interactions, they create logical
relationships among the information sources, 
not just the consumers and sharers of these sources.
But they are not like traditional page-to-page interactions either,
because they are not based on a hyperlink from $A$ to $B$, and
they are not in general determined by the authors of either $A$ or $B$;
it is the readers {\em deciding how $A$ and $B$ should be used
in discussion} who are determining the logical link between them.
In this sense, invoked interactions between $A$ and $B$ are
not directly under the control of the authors of $A$ and $B$;
they form a kind of ``revealed interpretation'' of $A$ and $B$
once they are released into social media.

The idea that replying to page $A$ with page $B$ can create a
semantically meaningful connection between $A$ and $B$ formed the
basis of an elegant technique due to Frigerri, Adamic, 
Eckles, and Cheng \cite{adamic-snopes} for identifying pages to
debunk widely circulated rumors.
Drawing on the fact that snopes.com is a heavily-used site for
evaluating Internet rumors,
they demonstrated that many instances of a pair $(B,A)$, where
$A$ is a page appearing in Facebook posts and $B$ is a page on 
snopes.com, serves as strong evidence that $B$ is providing
a judgment on the credibility of $A$.
In this way, scanning the replies to posts where $A$ occurs
for pages residing on snopes.com
provides an automated method for identifying a page $B$
that can help users evaluate the veracity of $A$.

Given that invoked interactions between pages $A$ and $B$ more generally
are voluminous, transcending any one particular domain or use case,
what do we learn if we consider the set of all such interactions
as a network in its own right, latent in a social media platform?

\xhdr{The present work: The structure of invocation graphs}
We are interested in understanding the global structure of 
the network of invoked interactions, and developing 
methods that can probe this structure, particularly for
questions related to political interaction.
Because our primary focus in this work will be at the
granularity of news and blogging {\em sources}
rather than the individual pages they produce,
we will consider this network at the level of domains:
using a large Twitter dataset covering all of 2016
up to the U.S. Presidential election,
we say that an {\em invoked interaction} from domain $B$ to 
domain $A$ occurs whenever a user replies to a tweet containing
a page from domain $A$ with a page from domain $B$.

We define the {\em invocation graph} on a set of domains of interest
as follows: for all pairs of domains $A$ and $B$ where there is
at least one invoked interaction from $B$ to $A$,
we include a weighted directed edge $(B,A)$ whose weight
is equal to the number of such invoked interactions.
Because we want to study how portions of the invocation graph
reflect aspects of political interaction, we choose the node
set (i.e. the domains of interest) using a preprocessing step
that only includes domains that were extensively retweeted
with the official Twitter accounts of Hillary Clinton and Donald Trump,
and then apply some further filtering heuristics that we describe
in the next section.

After this filtering, we have an invocation graph 
that reflects the ways in which
Twitter users interacted with one another by invoking content from different
politically relevant domains over the course of 2016.
Figure~\ref{fig:reply_graph} shows a few domains of such 
an invocation graph with some
sample interactions from our data. 
We can see these interactions can be supportive, like an
in-depth New York Times report on Russian information operations
shared in reply to an article
from The Guardian on the same subject, or adversarial, like a Breitbart article
on the New York Times' taxes shared in response to a Times article on Donald
Trump's taxes. These replies exhibit rich structure and give a sense for how
complex political interactions unfold on social media

We can thus return to some of the initial motivating questions
and ask how this interaction was structured relative to a
political spectrum containing support for Clinton on the left
and support for Trump on the right.
We do this using a political spectrum induced from the data,
rather than relying on external domain knowledge. There are a number
of ways to do this 
\cite{benkler-spectrum,flaxman-spectrum,gentzkow-media-slant}
that yield broadly consistent results, and we employ a method
of Benkler et al (described further in the next section)
that bases the spectrum on the relative frequency
of co-tweets with the Clinton and Trump Twitter accounts
\cite{benkler-spectrum}.

\xhdr{Embedding the Invocation Graph in a Political Spectrum}
Having thus located the domains on a political spectrum, we now
have an {\em embedded} version of the invocation graph: 
the nodes, representing domains,
are embedded in a one-dimensional spectrum, and the weighted directed edges
span pairs of points in this spectrum.
We can now ask how the edges are distributed across distances on
the spectrum, ranging from short-range interactions that connect
domains of similar political orientation to long-range interactions
that reach across sides.
In examining these questions, it is crucial to recall that 
these links among domains are not defined by hyperlinks in the
content on these domains, but instead via the replies made by
users when they invoke this content in discussions:
it is not that the domains are replying to each other, but that
they are being invoked in replies by users.
The data thus reflects choices made by the consumers of the content,
rather than by authors of the content.

In Section 3, we propose a set of methods to analyze how the edges
of the invocation graph span the underlying spectrum.
A core component of these methods is, for a domain $x$, to consider
its {\em out-link distribution} --- the distribution of ``landing points''
on the political spectrum for all links out of $x$.
As we move from left to right along the spectrum, tracking the 
out-link distributions of domains we encounter along the way, 
do the means of these distributions tend to move from left to right as well,
or do they tend to move inversely from right to left?
The former case would indicate a positive correlation in the locations of
the source and target of an invoked interaction $(B,A)$, suggesting links
are used to connect to similar sides of the political spectrum;
the latter case would correspond to a negative correlation and
hence connections across the spectrum,
with domains on the left being invoked to reply to domains on the right,
and vice versa.

It is not a priori obvious which type of correlation we should
expect to see; and as a reinforcement of this fact, we find that the nature
of the correlation actually inverts over the course of 2016 leading
up to the U.S. Presidential election.
In the early parts of 2016 we have a positive correlation, 
with politically similar domains being invoked to reply to each other;
but by the time we reach the months directly preceding the election,
this same correlation measure has become negative, indicating that
most of the linking is now crossing the spectrum.
We verify this effect using multiple measures, including one in 
which we compare the trends across the spectrum to what we'd observe
in a {\em randomly rewired} version of the embedded graph.

We also propose a set of methods to identify inherent {\em asymmetries}
in the patterns of linkage: do replies from left to right have the same
structure as replies from right to left?
Using our measures, we find strong asymmetries in the 2016 Twitter data,
with domains on the right side of the spectrum having a disproportionately
high rate of out-links in the invocation graph and 
domains on the left side of the spectrum having a disproportionately
high rate of in-links.
This right-to-left flow in the replies persists across the entire
time range, and is a key characteristic of the structure.

Since a recurring theme in our analyses is the way in which 
replies increasingly engaged opposite sides of the political spectrum
as 2016 went on, it is interesting to ask whether we see a similar
effect in a more traditional user-to-user interaction graph,
with nodes corresponding to users and directed edges to replies
from one user to another.
To explore this, we adapt the techniques
developed for the invocation graph to a user-to-user graph built from Reddit.
Specifically, we analyze a snapshot of Reddit's politics subreddit,
\subp, for the same period of 2016 up to the election;
we classify users by whether they had posted to the Clinton or Trump
subreddits, and then look at the rate of replies among different types
of users.  
We find that the trend on Reddit closely tracks the trend in the
invocation graph built
from Twitter, with increasing linkage between the two sides as
the election approached.

Overall, our methodology suggests that the invocation graph on domains, 
and its embedding into a one-dimensional spectrum,
captures important aspects of political interaction on social media ---
the tendency of users to interact by invoking links to authored content,
and the use of these interaction patterns to thus reveal relationships
among the content based on usage in everyday discussion.

\section{Basic Definitions}
\label{sec:basic}
We begin with a Twitter dataset containing aggregate-level information about
tweet-reply pairs. For each month from January to November 2016 (the US Presidential election was held on November 8, 2016), the dataset
consists of pairs of domains $x_1$ and $x_2$ along with an accompanying count,
the number of times a tweet containing a page from domain $x_1$ was posted in
reply to a tweet containing a page from domain $x_2$. In addition, for each
month, we have an auxiliary dataset of co-occurrences: for each domain $x$, the
number of times a user posted a tweet containing a page from $x$ on the same day
that the user retweeted Hillary Clinton's or Donald Trump's personal Twitter
account. Finally, we have the number of retweets of Clinton or Trump for each
month.

The invocation graph we construct from this data is a directed graph with domains as
vertices, where each domain corresponds to a news source. We draw an edge $x_1
\to x_2$ if a tweet containing a URL from domain $x_1$ is posted in reply to a
tweet containing a URL from domain $x_2$. The weight of this edge is the number
of such tweet-reply pairs. On first inspection, the most prominent feature of
this graph is that self-loops (consisting of links from
a domain $x$ to itself) have much higher weight than other edges. Since
our goal is to examine political interactions between domains, we remove all
self-loops from the graph.

\xhdr{Isolating Political Domains} The first issue we encounter is that Twitter
contains a wide range of URLs, not just pages from politically relevant domains. 
We could select only known political domains by whitelisting them, i.e. only
considering the subgraph over a predefined set of domains; however, this
approach will inevitably miss out on influential but less well-known news
sources.

On the other hand, there are challenges to a completely unsupervised approach.
URLs on Twitter are dominated by social media sites (e.g.\ twitter.com,
facebook.com) as well as content-hosting sites (e.g.\ imgur.com,
bitly.com) which produce virtually no content of their own, 
but instead host user-uploaded
content such as images, links, and text. While the usage of these
content-hosting sites would be interesting to study, this is outside the scope
of our work.

We begin by blacklisting several known social media and content-hosting domains and 
remove them from the graph. However, there are plenty of domains that appear
on Twitter that are not politically relevant, and we cannot individually remove
each such domain. To filter out such domains, we need some measure of
\emph{political engagement} for each domain. We can construct such a measure by
using the observation that politically relevant domains should frequently
co-occur with known political entities -- in our case,
the official Twitter accounts of Hillary Clinton and Donald Trump.

Our measure of political engagement for a domain, then, is simply the number of
times a user posted a tweet with that domain on the same day that he or she
retweeted either Clinton's or Trump's official Twitter account. Intuitively, the
more politically engaging a domain is, the more it will co-occur with these
political entities. With this proxy, we can select domains with high political
engagement, excluding social-media and content-hosting domains.

As a final filter, we require that each domain have an edge of at weight least
$W$ to some other domain in the political subgraph. This restricts our attention
to the most actively used political domains. Based on this, we can formally
define the invocation graph. Every domain in the invocation graph
\begin{itemize}
  \item is not blacklisted (social media and content-hosting
    domains)
  \item has political engagement above some threshold $p$
  \item has at least one edge to another domain in the invocation graph with weight
    at least $W$
\end{itemize}
We require the invocation graph be connected and contain some seed
domain. 
Any choice of popular American
news outlet would yield exactly the same result, as they all belong to the same
connected component.  (We use nytimes.com, but the choice is immaterial.)

Based on this definition, the algorithm to construct the invocation graph is as
follows: begin with the full graph, remove all nodes that are either blacklisted
or have political engagement below $p$, and run a breadth-first search beginning
at nytimes.com following only edges of weight at least $W$. In practice, we find
that the values we use for $p$ and $W$ don't affect our results much, and all
of the results presented here use $p = 10000$ and $W = 100$. Because we study
the change in this invocation graph over time, we build a new graph $G^m$ for each
month $m$ from January to November 2016.

\begin{figure}[ht]
  \centering
  \includegraphics[width=0.4\textwidth]{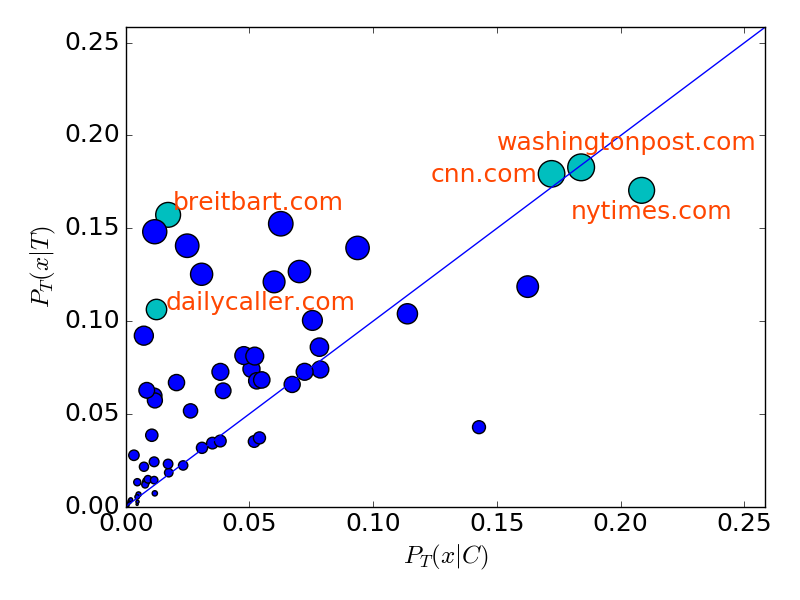}
  \caption{$\pxct$ vs. $\pxtt$ for September 2016}
  \label{fig:pxc_pxt_sep}
\end{figure}

\xhdr{A Political Spectrum} In order to characterize the political nature of
this graph, we need some way of organizing the domains along a political
spectrum. Drawing on techniques from \cite{benkler-spectrum} and our definition
of political engagement, we define the quantities $\pxct$ and $\pxtt$ for each
domain $x$ as the empirical probabilities that a user posted a tweet containing
a URL from domain $x$ given that, on the same day, he/she retweeted Clinton's or
Trump's official account respectively. Intuitively, if a user has retweeted Clinton on a
given day, the domains that he/she invokes are more likely to be on
Clinton's end of the political spectrum, and the same is
true for Trump. Figure~\ref{fig:pxc_pxt_sep} shows the resulting values, where
the blue line is $\pxct = \pxtt$. Interestingly, most domains lie above this
line, suggesting a difference in the populations of users 
retweeting Clinton and Trump respectively.

Furthermore, we can condense this information into a single \emph{political
score} for each domain:
\begin{equation}
  \pst{x} = \frac{\pxtt}{\pxct + \pxtt}
  \label{eq:pol_score}
\end{equation}
Note that $\pst{x} \in [0, 1]$, and the larger $\pst{x}$ is, the closer $x$ is to the
Trump end of the spectrum. Throughout this section, we use the spectrum built on
January-September 2016.

\section{Methodology}
\label{sec:method}
%!TEX root = ./main.tex

\begin{figure*}[ht]
  \centering
  \subfloat[January 2016]{%
    \includegraphics[width=0.4\textwidth]{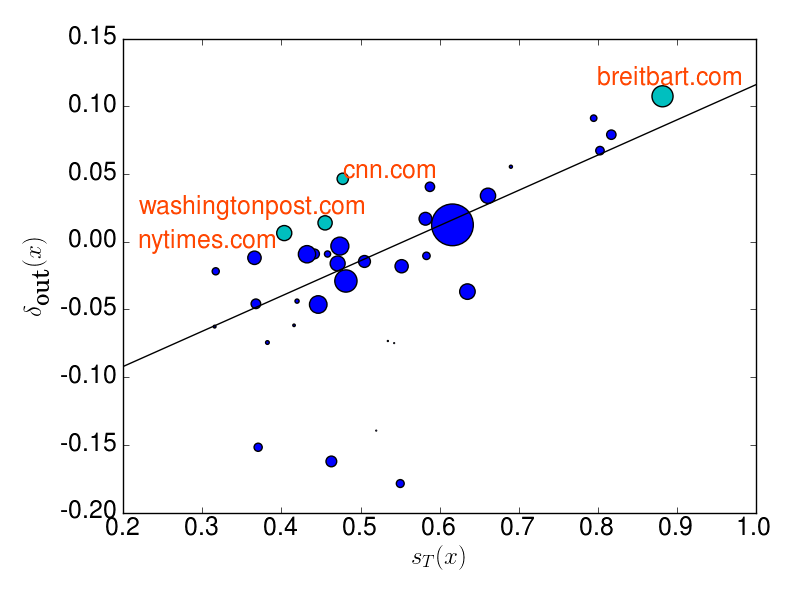}
    \label{fig:spec_outlink_jan}%
  }%
  \qquad \qquad
  \subfloat[October 2016]{%
    \includegraphics[width=0.4\textwidth]{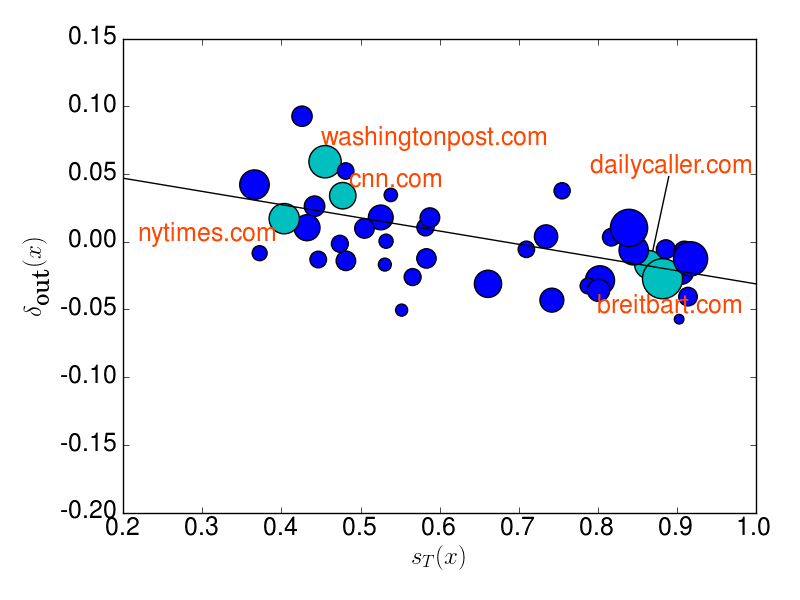}
    \label{fig:spec_outlink_oct}%
  }%
  \caption{Correlation between $\pst{x}$ and $\meanoutdiff(x)$ for
  January and October}
  \label{fig:spec_outlink}
\end{figure*}

With these definitions, we can analyze various properties of the invocation graph,
and in particular, its interaction with the political spectrum. Do usage
patterns of news articles and other political content differ across the
political spectrum? How do these patterns change leading up to the election?
Using the invocation graph in tandem with the political spectrum, we can shed light
on these questions.

\xhdr{Out-link Distributions} A key characteristic we study is the
\emph{out-link distribution} for a domain $x$: the distribution of positions on the
political spectrum where edges originating from $x$ land. A domain's
out-links describe the way in which it is used to reply to other domains. For
example, if a domain has many out-links to other domains near it on the
political spectrum, one might expect that it is being used to reinforce a
particular point of view; however, if it has many out-links to the opposite end
of the spectrum, it may instead be used to disagree or argue with an opposing
viewpoint.

To draw upon this intuition, we ask how $x$'s out-link distribution varies based
on its political score $\pst{x}$. As a baseline for comparison, we compute the
\emph{global} out-link distribution --- the distribution over the political
spectrum of where edges in $G$ land. Comparing to this baseline will give us some insight as
to how $x$'s linking pattern differs from the ``average'' linking pattern.

However, since $G$ contains no self-loops, $x$ doesn't link to itself, while the
global out-link distribution contains links to $x$. To prevent this from biasing
the comparison, we instead compare $x$'s out-link distribution to the out-link
distribution of $\gmx$, that is, $G$ with $x$ and all edges into or out of $x$
removed.

Thus, $x$'s out-link distribution distribution $d(x, G)$ is a distribution over
$[0, 1]$ assigning probability mass to $y \in [0,1]$ proportional to the weight
of edges from $x$ to $x'$ such that $\pst{x'} = y$. The global out-link
distribution $D(\gmx)$ can then be expressed as
\[
  D(\gmx) = \frac{\sum_{x' \in \gmx} d(x', \gmx) \cdot \vol(x', \gmx)}{\sum_{x'
  \in \gmx} \vol(x', \gmx)},
\]
where $\vol(x', \gmx)$ is the total weight of edges leaving $x'$ in $\gmx$. In
other words, $D(\gmx)$ is the weighted average of $d(x', \gmx)$ for $x' \in
\gmx$.

We make the comparison between these distributions formal as follows: for each
$x$, let $\meanout(x)$ be the weighted average political score of the domains
that $x$ links to, so that $\meanout(x) = \E{d(x, G)}$. This gives an estimate of
what types of domains $x$ is used to reply to --- if $\meanout(x)$ is close to 0,
then $x$ is used primarily to reply to Clinton-related domains, while if it is
close to 1, then $x$ is used primarily to reply to Trump-related domains. In a
slight abuse of notation, let $\meanout(\gmx)$ be the weighted average political
score of all endpoints of edges in $\gmx$, so $\meanout(\gmx) = \E{D(\gmx)}$.
Then,
\begin{equation}
  \meanoutdiff(x) = \meanout(x) - \meanout(\gmx)
  \label{eq:deltaout}
\end{equation}
is a measure of how much $x$ deviates from other domains towards the
Trump end of the political spectrum.

From past work documenting \emph{homophily} in online political
activity~\cite{adamic-blogosphere}, one might expect that a domain will be used
primarily to engage with politically similar domains, implying that
$\pst{x}$ is positively correlated with $\meanoutdiff(x)$. Furthermore, it is
not a priori obvious whether this trend should become more or less pronounced
leading up to the election --- do the opposing parties increasingly converse
within themselves, or do they engage with one another?

Figure~\ref{fig:spec_outlink} shows that in January, domains are more likely to
have links in the graph to politically similar domains, so $\pst{x}$ and
$\meanoutdiff(x)$ are positively correlated. However, by October, this
correlation has reversed: 
on average, domains are being used to reply \emph{across}
the political spectrum instead of to politically similar domains. Replies seem
to be adversarial, as with the breitbart.com $\to$ nytimes.com edge in
Figure~\ref{fig:reply_graph}.

\begin{figure}[ht]
  \centering
  \includegraphics[width=0.35\textwidth]{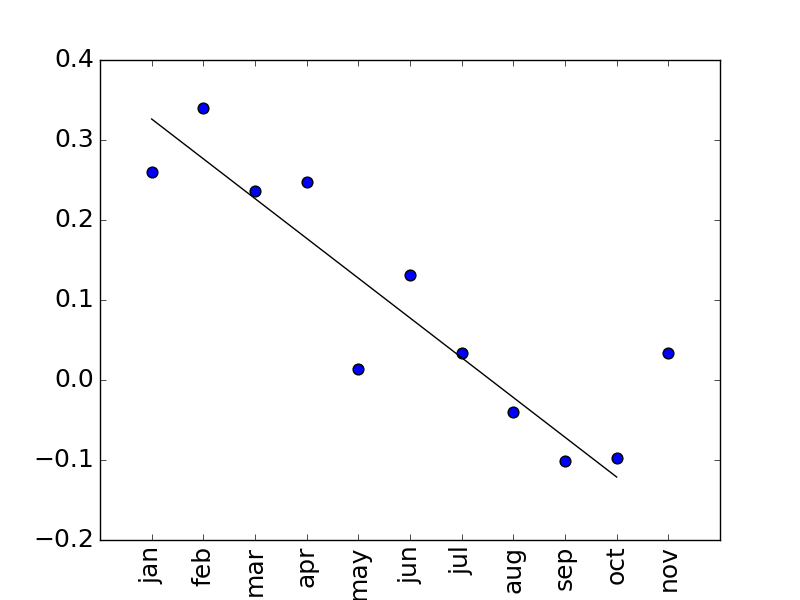}
  \caption{$a(G)$ for January-November}
  \label{fig:slopes_over_month}
\end{figure}

To understand this change in correlation over time, we define $a(G^m)$ to be the
slope of the correlation between $\pst{x}$ and
$\meanoutdiff(x)$ for month $m$. Figure~\ref{fig:slopes_over_month} shows that
$a(G^m)$ decreases significantly from January to October 2016. This suggests
that as the election drew nearer, the fraction of interaction between 
opposite ends of the spectrum
increased, and domains from the two opposing sides were 
actually being invoked more often to reply to each other.

\begin{figure}[ht]
  \centering
  \includegraphics[width=0.35\textwidth]{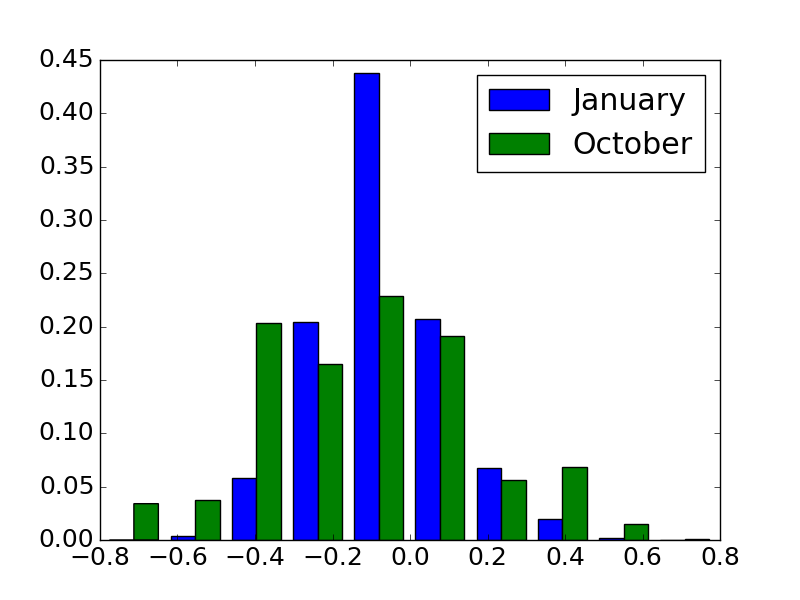}
  \caption{Edge length distributions for January vs. October}
  \label{fig:edge_lengths}
\end{figure}

\xhdr{Edge Lengths and Crossing Points} Another way of characterizing the
political aspects of the invocation graph is to consider the lengths and locations of
the edges on the political spectrum. 
As the interaction between opposing viewpoints
increases, we expect to see longer edges in the invocation graph that cross the
political spectrum instead of staying on one side. Figure~\ref{fig:edge_lengths}
shows that this trend holds true between January and October --- while most links
had length close to 0 in with respect to the political spectrum in January, by
October, many longer edges were present.

To enable us to visualize where on the spectrum edges lie, we make the following
definitions. For a point $y \in [0,1]$, let $\flr(y, G)$ be the number of edges
$(x_1, x_2)$ such that $\pst{x_1} < y < \pst{x_2}$. Similarly, let $\frl(y, G)$
be the number of edges such that $\pst{x_1} > y > \pst{x_2}$. In other words,
$\flr(y, G)$ is the number of edges crossing $y$ from left to right on the
political spectrum, and $\frl(y, G)$ is the number of edges crossing $y$ from
right to left.

\begin{figure*}[ht]
  \centering
  \subfloat[January 2016]{%
    \includegraphics[width=0.4\textwidth]{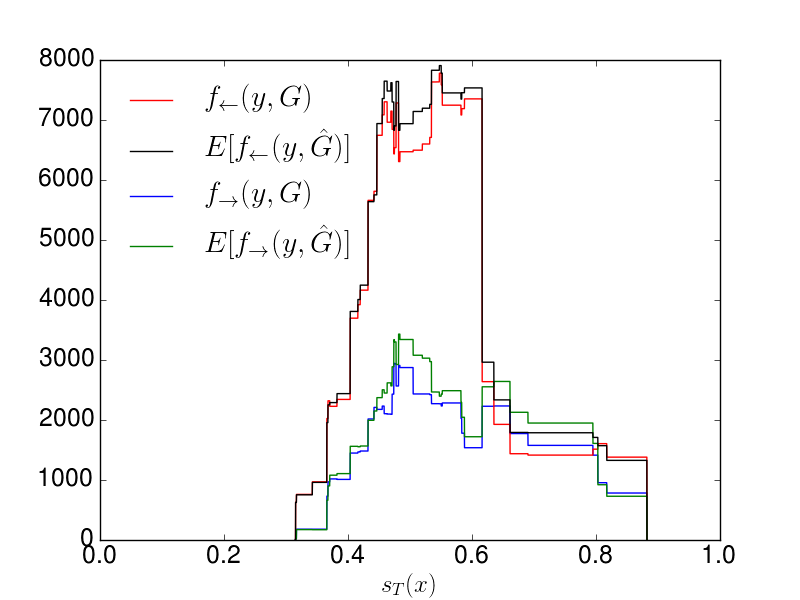}
    \label{fig:crossing_jan}%
  }%
  \qquad \qquad
  \subfloat[October 2016]{%
    \includegraphics[width=0.4\textwidth]{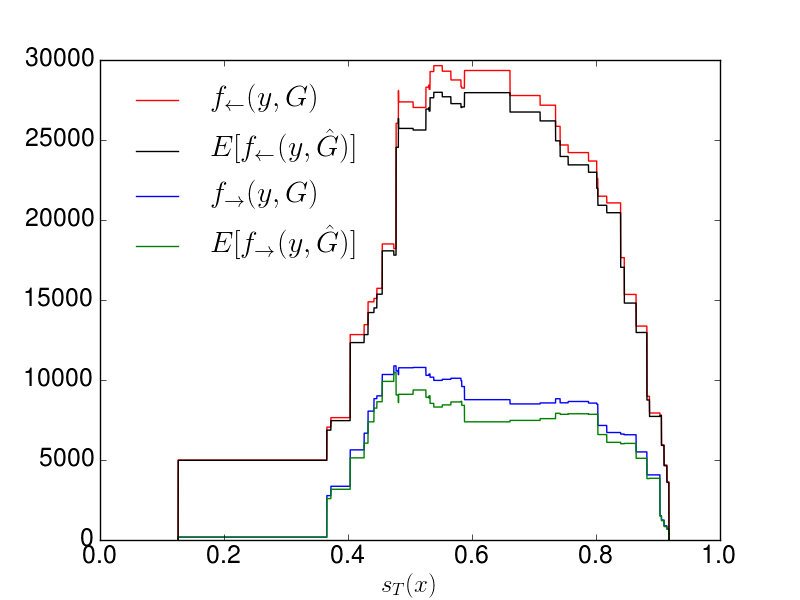}
    \label{fig:crossing_oct}%
  }%
  \caption{$\flr(y)$ and $\frl(y)$ for $G$ and $\rw{G}$}
  \label{fig:crossing}
\end{figure*}

In order to interpret these functions, we need a baseline to compare against. A
natural baseline in such scenarios is the \emph{randomly rewired} graph
$\rw{G}$, the idea of which goes back to~\cite{molloy1995critical}. In $\rw{G}$,
every vertex has the same indegree and outdegree as in $G$, but each edge has a
randomly chosen endpoint. Note that this can create
self-loops, which $G$ does not have by construction; however, the number of
self-loops in $\rw{G}$ is small (often 0 for a given randomization) and
therefore their effect on this analysis is negligible. Figure~\ref{fig:crossing}
shows $\flr$ and $\frl$ for $G$ and $\rw{G}$, where the values shown for
$\rw{G}$ are in expectation. In January, both $\flr$ and $\frl$ are dominated by
their rewired counterparts, while in October, the opposite is true. This
suggests that at the beginning of the year, domains were used to reply to
politically similar domains, resulting in shorter edges than a random baseline,
while closer to the election, domains were used to reply across the political
spectrum. This comparison allows us to get a sense for how actual behavior
deviates from random behavior, and how this deviation changes over time.

\begin{figure}[ht]
  \centering
  \includegraphics[width=0.4\textwidth]{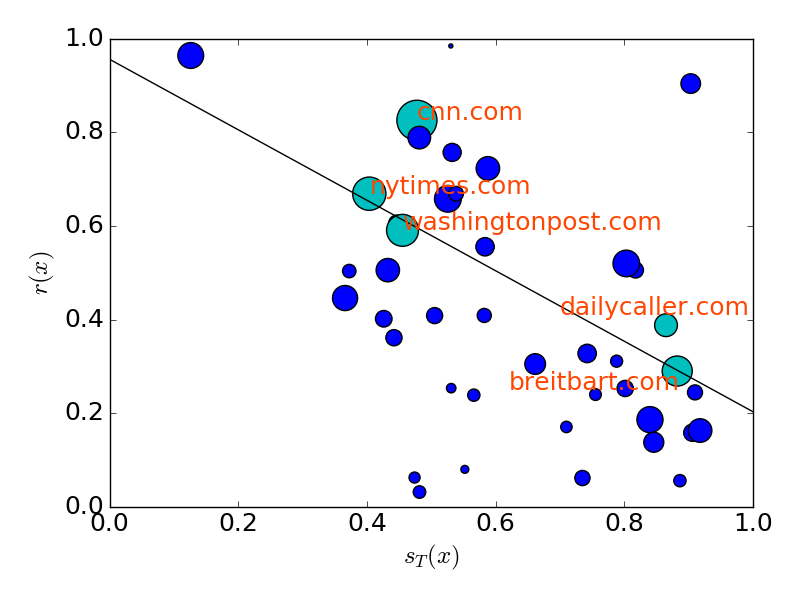}
  \caption{$\pst{x}$ vs. $r(x)$ for October}
  \label{fig:in_out}
\end{figure}

\xhdr{Asymmetry in Out-links} Another striking feature of
Figure~\ref{fig:crossing} is that $\frl$ dominates $\flr$, showing that many
more links crossed right-to-left than left-to-right. Intuitively, it seems that
a disproportionate number of edges originate on the right and end on the left,
corresponding to right-leaning domains being used to reply to left-leaning
domains. We can make this precise by defining $r(x)$ as the ratio of
$\text{indegree}(x) / (\text{indegree}(x) + \text{outdegree}(x))$ and analyzing
how $r(x)$ changes with $\pst{x}$. Figure~\ref{fig:in_out} shows that $r(x)$ is
negatively correlated with $\pst{x}$, meaning that domains on the right end of
the political spectrum produce a disproportionate number of out-links compared
to domains on the left end. In other words, domains on the right are more often
used to reply to other domains, while domains on the left are more often the
recipients of replies.

\section{Comparing to the User Level}
\label{sec:userlevel}
%!TEX root = ./main.tex
Having established a set of results for the structure of invocation graphs on
Twitter, we would like to verify that our findings are qualitatively consistent
with what we see in more traditional user-to-user communication graphs on social
media. Since our Twitter dataset doesn't contain information about individual
users, we instead turn to a publicly available Reddit
dataset\footnote{\url{https://files.pushshift.io/reddit/}}. Reddit is a
community discussion website organized into posts, or ``submissions,'' and
comments on those submissions. Comments are threaded, so that a comment is either
in reply to a top-level post or to another comment. The data consists of every
post and comment from Reddit in 2016, along with its author's username. Reddit
is subdivided into forums for particular topics called \emph{subreddits}. We
focus on three subreddits in particular: \subp, \subc, and \subt, which are
devoted to politics, Hillary Clinton, and Donald Trump respectively. All three
subreddits were among the most active subreddits during 2016. Note that in
addition to studying user-to-user dynamics on Reddit, in principle we could also
use Reddit data to replicate our Twitter invocation-graph analysis at the domain
level; however, it turns out that Reddit contains too few comments with URLs for
robust domain-level trends to emerge.

\xhdr{Interactions at the User Level} In order to test for analogous trends at the
user level to those we found at the domain level, we need to modify our methods.
In particular, we now need some kind of political information
about users. Whereas in Section~\ref{sec:basic} we anchored the political
spectrum to the official Clinton and Trump Twitter accounts, here we anchor our notion of political affiliations to \subc\ and \subt. Since most users are active in at most one
of the two subreddits, we have a simpler notion of a political score: we define
a set $\setc$ of users who posted in \subc\ but not \subt, and we define a set $\sett$ of users who posted in \subt\ but not \subc. There are 22,164 users in $\setc$ and 281,334 users in $\sett$ (more
than ten times the size of $\setc$). We assume that most users in
$\setc$ are pro-Clinton, while most users in $\sett$ are pro-Trump 
(this is consistent with the explicit ground rules
for participating in these subreddits).

\xhdr{Validating Political Information from Subreddits} We verify that
\subc\ and \subt\  contain strong signal about political orientation by 
adapting our methodology from Twitter to build a spectrum over the domains on
Reddit, and then comparing this spectrum to the one built from Twitter. To do this, we
define $\pxcr$ to be the empirical probability that a post or
comment in \subc\ contains a URL from domain $x$ (and analogously for $\pxtr$ and \subt). As
in~\eqref{eq:pol_score}, we can define a political score from Reddit as
\begin{equation}
  \psr{x} = \frac{\pxtr}{\pxcr + \pxtr}.
  \label{eq:reddit_s}
\end{equation}

Table~\ref{tab:1d_comp} shows the orderings of the Twitter and Reddit
spectra for 21 domains. The Spearman rank correlation~\cite{spearman1904proof} (a
measure of squared distance) between the two orderings is $0.871$ (compared
to a maximum of $0.757$ over $10,000$ randomly shuffled orderings). Thus the two settings align well, demonstrating that our notion of political
affiliation is adaptable to Reddit, and that it contains strong and consistent signal.

\begin{table}
\small
  \centering
  \begin{tabular}{|c|c|c|}
    \hline
    & Twitter & Reddit \\
    \hline
    1 & donaldjtrump.com & thegatewaypundit.com \\
    \hline
    2 & thegatewaypundit.com & zerohedge.com \\
    \hline
    3 & breitbart.com & breitbart.com \\
    \hline
    4 & dailycaller.com & donaldjtrump.com \\
    \hline
    5 & zerohedge.com & dailycaller.com \\
    \hline
    6 & foxnews.com & dailymail.co.uk \\
    \hline
    7 & nypost.com & foxnews.com \\
    \hline
    8 & dailymail.co.uk & nypost.com \\
    \hline
    9 & thehill.com & bbc.co.uk \\
    \hline
    10 & politico.com & theguardian.com \\
    \hline
    11 & cbsnews.com & cbsnews.com \\
    \hline
    12 & nbcnews.com & cnn.com \\
    \hline
    13 & cnn.com & thehill.com \\
    \hline
    14 & washingtonpost.com & nbcnews.com \\
    \hline
    15 & bbc.co.uk & huffingtonpost.com \\
    \hline
    16 & theguardian.com & washingtonpost.com \\
    \hline
    17 & nytimes.com & nytimes.com \\
    \hline
    18 & huffingtonpost.com & politico.com \\
    \hline
    19 & politifact.com & newsweek.com \\
    \hline
    20 & newsweek.com & politifact.com \\
    \hline
    21 & hillaryclinton.com & hillaryclinton.com \\
    \hline
  \end{tabular}
  \caption{Comparison of Twitter and Reddit political spectra}
  \label{tab:1d_comp}
  \vspace*{-0.2in}
\end{table}

\begin{figure*}[ht]
  \centering
  \subfloat[Comment counts]{%
    \includegraphics[width=0.3\textwidth]{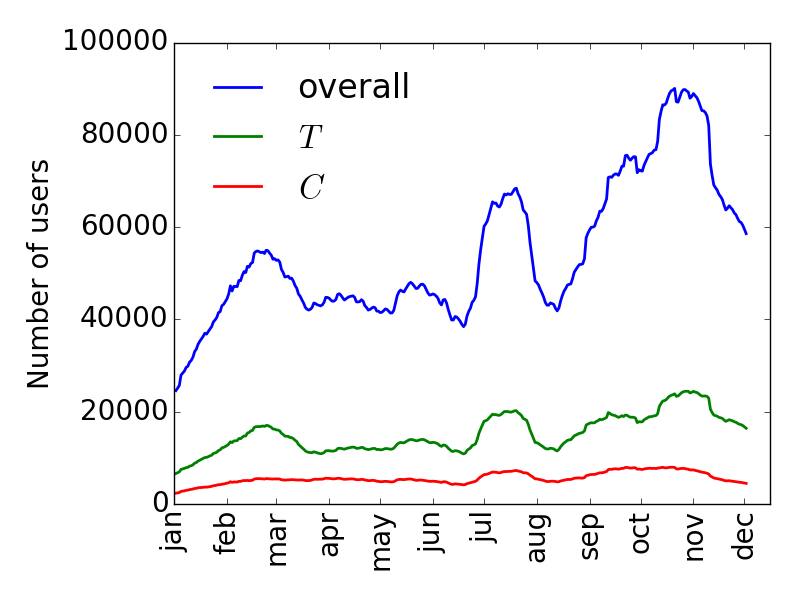}
    \label{fig:user_counts}
  }%
  \subfloat[Interaction types]{%
    \includegraphics[width=0.3\textwidth]{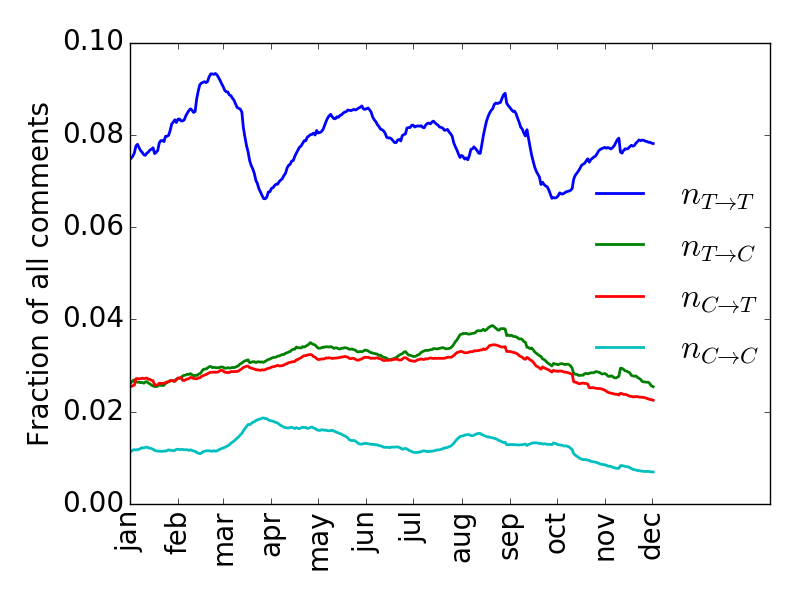}
    \label{fig:user_fracs}%
  }%
  \subfloat[Ratio of replies across the political spectrum]{%
    \includegraphics[width=0.3\textwidth]{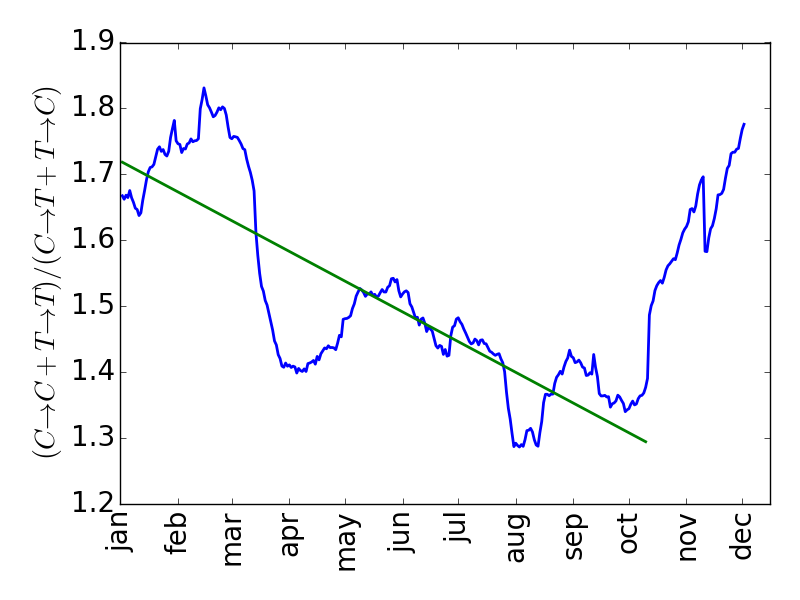}
    \label{fig:same_over_diff}%
  }%
  \caption{User-level trends from the Reddit dataset. In
  Figure~\ref{fig:same_over_diff}, a decrease in $(n_{C \to C} + n_{T \to
  T})/(n_{C \to T} + n_{T \to C})$ shows more interaction reaching across the
  spectrum.}
\end{figure*}

\xhdr{Building a User-to-User Graph on Reddit}
With this notion of political affiliation, we can now investigate some basic properties of political discourse on Reddit. Just as we first restricted our attention to political domains on Twitter, here we restrict our attention to the main political subreddit \subp. Since only a subset of the users in \subp\ are in
either $\setc$ or $\sett$, we focus on comment-reply pairs in which both users
involved are in one of $\setc$ or $\sett$. There are 4 possible types of interaction: $\setc \to \setc$, $\setc \to \sett$, $\sett \to \setc$, and
$\sett \to \sett$ where $p \to q$ means that a user from $p$ posts a reply to a
comment from a user in $q$. Let $n_{p \to q}$ be the number of $p \to q$
interactions in a given time period. We organize the data into sliding windows of
30 days. In the following plots, the value at a particular date represents the
30-day window ending at that date. Figure~\ref{fig:user_counts} shows comment
counts throughout 2016 averaged over the
next 30 days, and Figure~\ref{fig:user_fracs} shows what fraction of the
comments were each type of interaction.

Figure~\ref{fig:user_fracs} shows that a steadily rising
number of comments are cross-cutting (e.g., are of types $\setc \to \sett$ and $\sett \to \setc$) from the beginning of 2016 up until the election in early November, followed by a return to a baseline rate. Figure~\ref{fig:same_over_diff} further reinforces this point, showing a strong negative slope in the ratio of edges
between users of the same political leaning and
users of different political leanings from
January until the November 8 election. These results are consistent with our findings our Twitter, where an increasing fraction of political interactions reach across the political spectrum on Twitter up until the election. 

To interpret the significance of these results, we adapt our random
rewiring technique to the user level by randomly reassigning users to comments,
preserving
the invariant that each user still posts the same number of
comments as in the original data. Whether we randomize globally (a comment is
randomly attributed to any user from all of 2016) or each month
(a comment is attributed to a randomly chosen user from the same month that
comment was written), the results are the same -- the observed slope is
significantly more negative than the minimum over 100 random trials.

Using our Reddit dataset, we've shown that our invocation graph methodology can be
adapted to analyze traditional user-user communication graphs. In particular, we've used co-occurrences to determine political
information about both users and domains. The month-to-month trend found at the
user level mirrors our findings on Twitter --- in the months leading up to the
election, online political interaction increasingly reached across the political spectrum.

\section{Two-Dimensional Alignment}
%!TEX root = ./main.tex
Our formulation of the political spectrum in this work
indicates that it has a natural
{\em two-dimensional} structure, with one dimension corresponding
to co-occurrence probabilities with content related to one candidate,
and the other dimension
corresponding to co-occurrence probabilities 
with content related to the other candidate.

In Section~\ref{sec:userlevel}, when we established that our invocation
graph methodology extends naturally to the traditional user-level
setting, part of our analysis involved measuring how well the
one-dimensional spectra $\pst{\cdot}$ and $\psr{\cdot}$ shown in
Table~\ref{tab:1d_comp} align. 
Here we consider how to measure the alignment of the corresponding
two-dimensional spectra.

To make this comparison, we need to account for the fact
that axes may have different scales (e.g. posts on Reddit
contain URLs at a different rate than on Twitter). This means we
need to scale $\pxcr$ and $\pxtr$ in order to find the ``best match'' between
the two spectra. We formalize this as the following optimization problem,
minimizing the squared $\ell_2$ distance of each pair of points:
\begin{equation}
  \min_{a,b: a,b \ge 0} \sum_{x \in D} (\pxct - a\pxcr)^2 + (\pxtt - b
  \pxtr)^2
  \label{eq:opt_l2}
\end{equation}
where $D$ is the set of domains. Since all points lie in the first quadrant, we
can drop the constraint $a,b \ge 0$. Note that~\eqref{eq:opt_l2} can be
separated into 2 identical optimization problems of the form
\begin{equation}
  \min_c \sum_{x \in D} (u_x - cv_x)^2.
  \label{eq:opt_l2_simple}
\end{equation}
If $u$ and $v$ are the vectors $[u_x : x \in D]\tran$ and $[v_x : x \in D]^\top$
respectively, then this can be written as $\min_c \sum_{x \in D} \|u - cv\|_2^2$.
This is convex and has derivative
\[
  \frac{d}{dc} \|u - cv\|_2^2 = \frac{d}{dc} (u-cv)\tran (u-cv) = 2v\tran u -
  2cv\tran v.
\]
Setting this equal to 0, we find that~\eqref{eq:opt_l2_simple} has the solution
$c = \frac{v\tran u}{v \tran v}$.

\begin{figure}[ht]
  \centering
  \includegraphics[width=.4\textwidth]{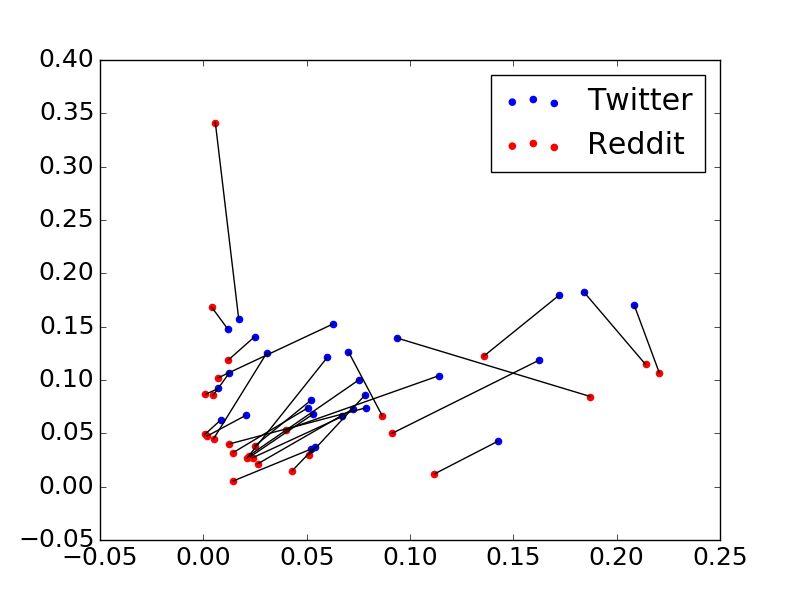}
  \caption{Scaled spectrum comparison}
  \label{fig:2d_comp}
\end{figure}

Using this, we can scale the Reddit spectrum to compare it to the Twitter
spectrum, producing the plot shown in Figure~\ref{fig:2d_comp}. The spectra
roughly align, with the domains in approximately the same positions 
for both Twitter and Reddit.

If instead we wish to minimize $\ell_1$ distance between pairs of points, we get
the optimization problem
\begin{equation}
  \min_{a,b: a,b \ge 0} \sum_{x \in D} |\pxct - a\pxcr| + |\pxtt - b
  \pxtr|
  \label{eq:opt_l1}
\end{equation}
Again, this results in two separate optimizations of the form
\begin{equation}
  \min_c \sum_{x \in D} |u_x - cv_x|.
  \label{eq:opt_l1_simple}
\end{equation}
The subgradient of each term is
\[
  \begin{cases}
    \{v_x \cdot \text{sign}(cv_x - u_x)\} & u_x \ne cv_x \\
    [-v_x, v_x] & \text{otherwise}
  \end{cases}
\]
We sort the $x$'s by increasing $u_x/v_x$, so $i < j \Longrightarrow
u_{x_i}/v_{x_i} \le u_{x_j}/v_{x_j}$, and then choose $i$ such that
\[
  \sum_{j \le i} v_{x_j} > \sum_{j > i} v_{x_j} ~~~
  \text{and} ~~~
  \sum_{j < i} v_{x_j} < \sum_{j \ge i} v_{x_j}.
\]
Essentially, $i$ \emph{balances} the $v_{x_j}$'s, splitting them into two
sets such that adding $v_{x_i}$ to either set gives that set the larger sum
of the two. Thus, the subgradient of~\eqref{eq:opt_l1_simple} contains 0
for $c = \frac{u_{x_i}}{v_{x_i}}$, making this the solution to~\eqref{eq:opt_l1}.

\begin{figure*}[ht]
  \centering
  \subfloat[January 2016]{
    \includegraphics[width=.3\textwidth]{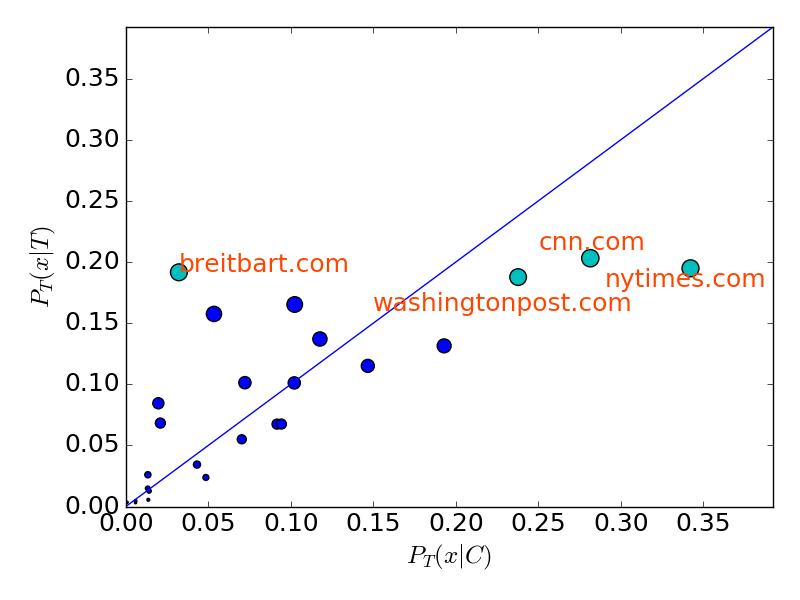}
    \label{fig:pxc_pxt_jan}
  }
  \subfloat[February 2016]{
    \includegraphics[width=.3\textwidth]{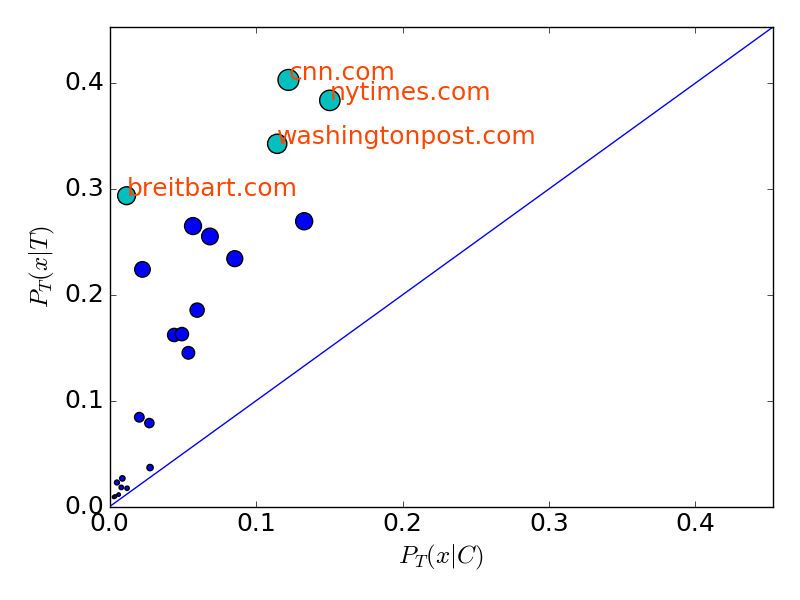}
    \label{fig:pxc_pxt_feb}
  }
  \subfloat[Scaled Comparison]{
    \includegraphics[width=.3\textwidth]{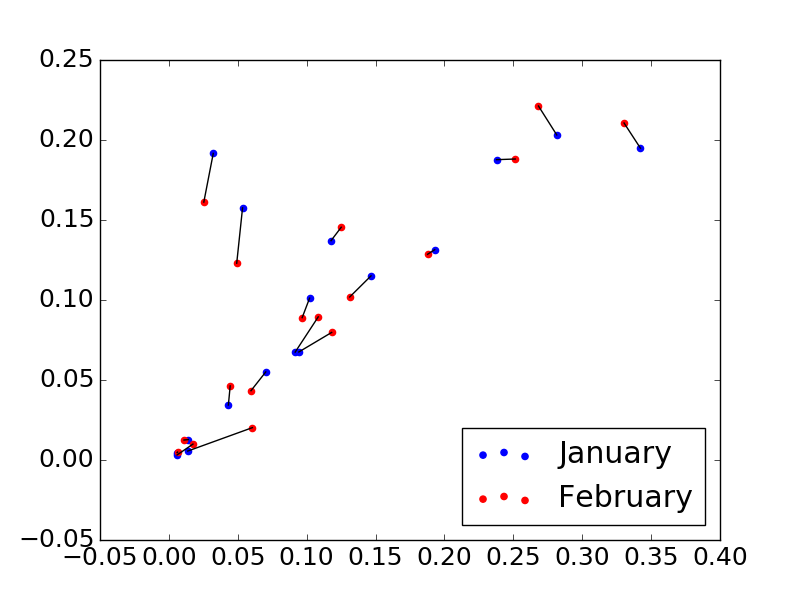}
    \label{fig:jan_feb_comp}
  }
  \caption{Spectrum Comparison for January and February 2016 on Twitter}
  \label{fig:jan_feb_scale}
\end{figure*}

We can also use this same alignment technique to compare spectra between
months. Figures~\ref{fig:pxc_pxt_jan} and~\ref{fig:pxc_pxt_feb}
show that in absolute terms, the spectra produced by January and February on
Twitter look quite different. In fact, after scaling the axes, the spectra
align very well (Figure~\ref{fig:jan_feb_comp}, using $\ell_1$-minimization).
One way to see this is to compare the quality of alignment in our data the
quality of alignment among shuffled versions of the spectra for
January and February, where the domains labeling the points are randomly
permuted; we find that the alignment for the real data has much
lower cost than the typical alignment for shuffled data.
The disparity between the unscaled spectra seems to indicate an influx of users
in February who retweeted Clinton but didn't necessarily engage in other sorts
of political activity; this left the relative position of other domains in the two-dimensional spectrum unaffected.

We also note this type of scaling doesn't affect the one-dimensional
orderings in Table~\ref{tab:1d_comp}. 
We can rewrite the political score as $\pst{x} = \frac{1}{1 + \tan \theta_T}$, 
where $\theta_T$ is the angle of $x$ from the $\pxct$ axis. Since scaling the
axes preserves the ordering of these angles, it also preserves the
one-dimensional rankings.

\section{Further Related Work}
%!TEX root = ./main.tex
Our work builds on rich literatures in online social media, online news, information diffusion, and the landscape of online political interaction, and it also draws on a long history of work studying the role of media in politics. 

With access to information now abundant, a growing line of work has investigated
how this impacts consumption of political information. 
As mentioned in the introduction, a key issue has been measuring the extent to which online political interaction crosses ideological divides, or whether it stays relatively confined in ``filter bubbles'' or ``echo chambers''~\cite{pariser-filter-bubble,sunstein2001echo,garrett2009echo,gilbert2009blogs,bakshy-ideologically-diverse,flaxman-spectrum}. 
Concerns about the Web potentially catalyzing balkanization, or fragmentation into isolated ideological divisions, date from the 1990s~\cite{van1996could}. 
Early empirical work on the structure of online political interaction
through blogging 
identified strong ideological partitions evident in large-scale analysis
of linking patterns~\cite{adamic-blogosphere}, whereas analyses
of more recent social media platforms
finds evidence for a more complex structure, in which both ideological
entrenchment and exposure to more diverse content is promoted on the
Web~\cite{flaxman-spectrum,bakshy-ideologically-diverse}.

More broadly, our paper relates to the extensive work on the structure of information sharing on the Web~\cite{bakshy-socnet-info-diffuse}, 
as well as theoretical work on how network structure affects information flow~\cite{jackson2006diffusion,golub2009homophily}.
This structure has been quantified in many ways, particularly by organizing shares into information cascades~\cite{goel-structural-virality,cheng-cascade-prediction}, where nodes are typically people and edges between people indicate if one person directly shared a piece of content with another person.
In contrast, our work here introduces and studies invocation graphs, where nodes are
information sources and edges indicate that a user shared
content from one source in response to content from another source.

The role of the media in politics is the subject of an active field of
study~\cite{graber2017mass,street2010mass}. Particularly relevant to our
work here is the study of how news and public opinion spread through
social networks, including the early theory of two-step
flow~\cite{katz1957two}. More recently, there has been a spirited
debate about the impact that ``influencers'' have on these
processes~\cite{watts2007influentials,cha2010measuring,bakshy2011everyone,katz1966personal,gladwell2006tipping}.

\section{Conclusion}
%!TEX root = ./main.tex
In this work we have introduced \emph{invocation graphs}, together with
a set of techniques for analyzing them, 
as a means of probing the structure of online
political interaction. In combination with previous methods for
inducing a political spectrum from data, we develop methods for
measuring several important phenomena. In particular, we analyzed 
a natural embedding of the invocation graph in the political spectrum, 
and asked how its edges are distributed across this 
spectrum---whether they are sequestered in ideological
pockets or whether they span larger ideological distances. Applying these
techniques to political interaction on Twitter in the months leading up to the 
2016 US Presidential election, we observed that political interaction
via the invocation graph 
became increasingly cross-cutting as the election neared. We also
developed methods to analyze whether there are inherent asymmetries
between how the right and the left engage each other via replies. 
Applying our techniques to Twitter, we found that edges 
in in the invocation graph more
consistently went from sources on the right to sources on the left
than in the other direction.

It is worth emphasizing a critical feature of invocation graphs, which is
that they are composed of \emph{invoked interactions} as opposed to
direct interactions. Although on a surface level invocation graphs may
resemble the hyperlink graphs that form a basic staple of Web analysis,
they are actually quite different, as the links are generally not
under the nodes' control. News sources publish content, and
then it is up to the readership to determine how these sources
connect to each other in the invocation graph. 
In this sense, relative positions and functions of news
domains in invocation graphs are indicative of how the public actually uses
them in online political discussion. Beyond the ideological
territory news sources may try to explicitly claim for themselves,
invocation graphs position these sources according to their roles in political
interaction.

There are a number of important directions that remain to be pursued.
First, it is intriguing to see how interaction both in the invocation graph
on Twitter and the user-to-user interaction graph on Reddit 
became more and more ideologically cross-cutting as the election
approached, given that homophily would suggest that most links
should link to nearby points on the spectrum.
Is there a systematic way to relate this trend to an underlying
level of polarization, so that changes in the structure of the
embedded invocation graph might provide insight into polarization and
how it evolves?
Second, beyond the two large social media datasets we considered,
it would be illuminating to apply our methods in other settings as well. 
In particular,
does online debate increasingly cross ideological divides in the
run-up to milestone events in general, or was this specific to the
2016 US Presidential election?   
In general, we believe that applying and extending our methods for
online political interaction using invocation graphs contains many promising
directions for future work.

\xhdr{Acknowledgements} 
We thank Lada Adamic, Glenn Altschuler, Isabel Kloumann,
and Michael Macy for valuable discussions about these topics.
MR is supported by an NSF Graduate Research Fellowship
(DGE-1650441).  JK is supported in part by 
a Simons Investigator Award, an ARO MURI grant, and NSF grant 1741441.
This work was performed in part while AA and MR were at Microsoft Research.

\bibliographystyle{ACM-Reference-Format}
\bibliography{refs}


\begin{thebibliography}{33}

%%% the Web address.

\ifx \showCODEN    \undefined \def \showCODEN     #1{\unskip}     \fi
\ifx \showDOI      \undefined \def \showDOI       #1{#1}\fi
\ifx \showISBNx    \undefined \def \showISBNx     #1{\unskip}     \fi
\ifx \showISBNxiii \undefined \def \showISBNxiii  #1{\unskip}     \fi
\ifx \showISSN     \undefined \def \showISSN      #1{\unskip}     \fi
\ifx \showLCCN     \undefined \def \showLCCN      #1{\unskip}     \fi
\ifx \shownote     \undefined \def \shownote      #1{#1}          \fi
\ifx \showarticletitle \undefined \def \showarticletitle #1{#1}   \fi
\ifx \showURL      \undefined \def \showURL       {\relax}        \fi
% invisible to TeX
\providecommand\bibfield[2]{#2}
\providecommand\bibinfo[2]{#2}
\providecommand\natexlab[1]{#1}
\providecommand\showeprint[2][]{arXiv:#2}

\bibitem[\protect\citeauthoryear{Adamic and Glance}{Adamic and Glance}{2005}]%
        {adamic-blogosphere}
\bibfield{author}{\bibinfo{person}{Lada~A. Adamic} {and}
  \bibinfo{person}{Natalie Glance}.} \bibinfo{year}{2005}\natexlab{}.
\newblock \showarticletitle{The Political Blogosphere and the 2004 U.S.
  Election: Divided They Blog}. In \bibinfo{booktitle}{\emph{Proceedings of the
  3rd International Workshop on Link Discovery}}
  \emph{(\bibinfo{series}{LinkKDD '05})}. \bibinfo{publisher}{ACM},
  \bibinfo{address}{New York, NY, USA}, \bibinfo{pages}{36--43}.
\newblock
\showISBNx{1-59593-215-1}
\urldef\tempurl%
\url{https://doi.org/10.1145/1134271.1134277}
\showDOI{\tempurl}


\bibitem[\protect\citeauthoryear{Bakshy, Hofman, Mason, and Watts}{Bakshy
  et~al\mbox{.}}{2011}]%
        {bakshy2011everyone}
\bibfield{author}{\bibinfo{person}{Eytan Bakshy}, \bibinfo{person}{Jake~M
  Hofman}, \bibinfo{person}{Winter~A Mason}, {and} \bibinfo{person}{Duncan~J
  Watts}.} \bibinfo{year}{2011}\natexlab{}.
\newblock \showarticletitle{Everyone's an influencer: quantifying influence on
  twitter}. In \bibinfo{booktitle}{\emph{Proceedings of the fourth ACM
  international conference on Web search and data mining}}. ACM,
  \bibinfo{pages}{65--74}.
\newblock


\bibitem[\protect\citeauthoryear{Bakshy, Messing, and Adamic}{Bakshy
  et~al\mbox{.}}{2015}]%
        {bakshy-ideologically-diverse}
\bibfield{author}{\bibinfo{person}{Eytan Bakshy}, \bibinfo{person}{Solomon
  Messing}, {and} \bibinfo{person}{Lada~A Adamic}.}
  \bibinfo{year}{2015}\natexlab{}.
\newblock \showarticletitle{Exposure to ideologically diverse news and opinion
  on Facebook}.
\newblock \bibinfo{journal}{\emph{Science}} \bibinfo{volume}{348},
  \bibinfo{number}{6239} (\bibinfo{year}{2015}), \bibinfo{pages}{1130--1132}.
\newblock


\bibitem[\protect\citeauthoryear{Bakshy, Rosenn, Marlow, and Adamic}{Bakshy
  et~al\mbox{.}}{2012}]%
        {bakshy-socnet-info-diffuse}
\bibfield{author}{\bibinfo{person}{Eytan Bakshy}, \bibinfo{person}{Itamar
  Rosenn}, \bibinfo{person}{Cameron~A. Marlow}, {and} \bibinfo{person}{Lada~A.
  Adamic}.} \bibinfo{year}{2012}\natexlab{}.
\newblock \showarticletitle{The Role of Social Networks in Information
  Diffusion}. In \bibinfo{booktitle}{\emph{Proc. World Wide Web Conference}}.
\newblock


\bibitem[\protect\citeauthoryear{Benkler, Faris, Roberts, and
  Zuckerman}{Benkler et~al\mbox{.}}{2017}]%
        {benkler-spectrum}
\bibfield{author}{\bibinfo{person}{Yochai Benkler}, \bibinfo{person}{Robert
  Faris}, \bibinfo{person}{Hal Roberts}, {and} \bibinfo{person}{Ethan
  Zuckerman}.} \bibinfo{year}{2017}\natexlab{}.
\newblock \showarticletitle{Study: Breitbart-led right-wing media ecosystem
  altered broader media agenda}.
\newblock \bibinfo{journal}{\emph{Columbia Journalism Review}}
  \bibinfo{volume}{1}, \bibinfo{number}{4.1} (\bibinfo{year}{2017}),
  \bibinfo{pages}{7}.
\newblock


\bibitem[\protect\citeauthoryear{Bennett}{Bennett}{1996}]%
        {bennett-news-illusion}
\bibfield{author}{\bibinfo{person}{W.~Lance Bennett}.}
  \bibinfo{year}{1996}\natexlab{}.
\newblock \bibinfo{booktitle}{\emph{News: The politics of illusion}}.
\newblock \bibinfo{publisher}{Longman}.
\newblock


\bibitem[\protect\citeauthoryear{Cha, Haddadi, Benevenuto, and Gummadi}{Cha
  et~al\mbox{.}}{2010}]%
        {cha2010measuring}
\bibfield{author}{\bibinfo{person}{Meeyoung Cha}, \bibinfo{person}{Hamed
  Haddadi}, \bibinfo{person}{Fabricio Benevenuto}, {and}
  \bibinfo{person}{P~Krishna Gummadi}.} \bibinfo{year}{2010}\natexlab{}.
\newblock \showarticletitle{Measuring user influence in twitter: The million
  follower fallacy.}
\newblock \bibinfo{journal}{\emph{ICWSM}} \bibinfo{volume}{10},
  \bibinfo{number}{10-17} (\bibinfo{year}{2010}), \bibinfo{pages}{30}.
\newblock


\bibitem[\protect\citeauthoryear{Cheng, Adamic, Dow, Kleinberg, and
  Leskovec}{Cheng et~al\mbox{.}}{2014}]%
        {cheng-cascade-prediction}
\bibfield{author}{\bibinfo{person}{Justin Cheng}, \bibinfo{person}{Lada~A.
  Adamic}, \bibinfo{person}{P.~Alex Dow}, \bibinfo{person}{Jon~M. Kleinberg},
  {and} \bibinfo{person}{Jure Leskovec}.} \bibinfo{year}{2014}\natexlab{}.
\newblock \showarticletitle{Can cascades be predicted?}. In
  \bibinfo{booktitle}{\emph{23rd International World Wide Web Conference, {WWW}
  '14, Seoul, Republic of Korea, April 7-11, 2014}}. \bibinfo{pages}{925--936}.
\newblock
\urldef\tempurl%
\url{https://doi.org/10.1145/2566486.2567997}
\showDOI{\tempurl}


\bibitem[\protect\citeauthoryear{Conover, Ratkiewicz, Francisco,
  Gon{\c{c}}alves, Menczer, and Flammini}{Conover et~al\mbox{.}}{2011}]%
        {conover-polarization-twitter}
\bibfield{author}{\bibinfo{person}{Michael Conover}, \bibinfo{person}{Jacob
  Ratkiewicz}, \bibinfo{person}{Matthew~R. Francisco}, \bibinfo{person}{Bruno
  Gon{\c{c}}alves}, \bibinfo{person}{Filippo Menczer}, {and}
  \bibinfo{person}{Alessandro Flammini}.} \bibinfo{year}{2011}\natexlab{}.
\newblock \showarticletitle{Political Polarization on Twitter}. In
  \bibinfo{booktitle}{\emph{Proceedings of the Fifth International Conference
  on Weblogs and Social Media, Barcelona, Catalonia, Spain, July 17-21, 2011}}.
\newblock
\urldef\tempurl%
\url{http://www.aaai.org/ocs/index.php/ICWSM/ICWSM11/paper/view/2847}
\showURL{%
\tempurl}


\bibitem[\protect\citeauthoryear{Dow, Adamic, and Friggeri}{Dow
  et~al\mbox{.}}{2013}]%
        {dow-facebook-cascades}
\bibfield{author}{\bibinfo{person}{P.~Alex Dow}, \bibinfo{person}{Lada~A.
  Adamic}, {and} \bibinfo{person}{Adrien Friggeri}.}
  \bibinfo{year}{2013}\natexlab{}.
\newblock \showarticletitle{The Anatomy of Large Facebook Cascades}. In
  \bibinfo{booktitle}{\emph{Proceedings of the Seventh International Conference
  on Weblogs and Social Media, {ICWSM} 2013, Cambridge, Massachusetts, USA,
  July 8-11, 2013.}}
\newblock
\urldef\tempurl%
\url{http://www.aaai.org/ocs/index.php/ICWSM/ICWSM13/paper/view/6123}
\showURL{%
\tempurl}


\bibitem[\protect\citeauthoryear{Flaxman, Goel, and Rao}{Flaxman
  et~al\mbox{.}}{2016}]%
        {flaxman-spectrum}
\bibfield{author}{\bibinfo{person}{Seth Flaxman}, \bibinfo{person}{Sharad
  Goel}, {and} \bibinfo{person}{Justin Rao}.} \bibinfo{year}{2016}\natexlab{}.
\newblock \showarticletitle{Filter Bubbles, Echo Chambers, and Online News
  Consumption}.
\newblock \bibinfo{journal}{\emph{Public Opinion Quarterly}}
  \bibinfo{volume}{80} (\bibinfo{year}{2016}).
\newblock


\bibitem[\protect\citeauthoryear{Friggeri, Adamic, Eckles, and Cheng}{Friggeri
  et~al\mbox{.}}{2014}]%
        {adamic-snopes}
\bibfield{author}{\bibinfo{person}{Adrien Friggeri}, \bibinfo{person}{Lada~A
  Adamic}, \bibinfo{person}{Dean Eckles}, {and} \bibinfo{person}{Justin
  Cheng}.} \bibinfo{year}{2014}\natexlab{}.
\newblock \showarticletitle{Rumor Cascades}. In
  \bibinfo{booktitle}{\emph{ICWSM}}.
\newblock


\bibitem[\protect\citeauthoryear{Garrett}{Garrett}{2009}]%
        {garrett2009echo}
\bibfield{author}{\bibinfo{person}{R~Kelly Garrett}.}
  \bibinfo{year}{2009}\natexlab{}.
\newblock \showarticletitle{Echo chambers online?: Politically motivated
  selective exposure among Internet news users}.
\newblock \bibinfo{journal}{\emph{Journal of Computer-Mediated Communication}}
  \bibinfo{volume}{14}, \bibinfo{number}{2} (\bibinfo{year}{2009}),
  \bibinfo{pages}{265--285}.
\newblock


\bibitem[\protect\citeauthoryear{Gentzkow and Shapiro}{Gentzkow and
  Shapiro}{2010}]%
        {gentzkow-media-slant}
\bibfield{author}{\bibinfo{person}{Matthew Gentzkow} {and}
  \bibinfo{person}{Jesse~M Shapiro}.} \bibinfo{year}{2010}\natexlab{}.
\newblock \showarticletitle{What drives media slant? {Evidence} from {US} daily
  newspapers}.
\newblock \bibinfo{journal}{\emph{Econometrica}} \bibinfo{volume}{78},
  \bibinfo{number}{1} (\bibinfo{year}{2010}), \bibinfo{pages}{35--71}.
\newblock


\bibitem[\protect\citeauthoryear{Gilbert, Bergstrom, and Karahalios}{Gilbert
  et~al\mbox{.}}{2009}]%
        {gilbert2009blogs}
\bibfield{author}{\bibinfo{person}{Eric Gilbert}, \bibinfo{person}{Tony
  Bergstrom}, {and} \bibinfo{person}{Karrie Karahalios}.}
  \bibinfo{year}{2009}\natexlab{}.
\newblock \showarticletitle{Blogs are echo chambers: Blogs are echo chambers}.
  In \bibinfo{booktitle}{\emph{HICSS'09. 42nd Hawaii International Conference
  on System Sciences}}. IEEE, \bibinfo{pages}{1--10}.
\newblock


\bibitem[\protect\citeauthoryear{Gladwell}{Gladwell}{2006}]%
        {gladwell2006tipping}
\bibfield{author}{\bibinfo{person}{Malcolm Gladwell}.}
  \bibinfo{year}{2006}\natexlab{}.
\newblock \bibinfo{booktitle}{\emph{The tipping point: How little things can
  make a big difference}}.
\newblock \bibinfo{publisher}{Little, Brown}.
\newblock


\bibitem[\protect\citeauthoryear{Goel, Anderson, Hofman, and Watts}{Goel
  et~al\mbox{.}}{2016}]%
        {goel-structural-virality}
\bibfield{author}{\bibinfo{person}{Sharad Goel}, \bibinfo{person}{Ashton
  Anderson}, \bibinfo{person}{Jake~M. Hofman}, {and} \bibinfo{person}{Duncan~J.
  Watts}.} \bibinfo{year}{2016}\natexlab{}.
\newblock \showarticletitle{The Structural Virality of Online Diffusion}.
\newblock \bibinfo{journal}{\emph{Management Science}} \bibinfo{volume}{62},
  \bibinfo{number}{1} (\bibinfo{year}{2016}), \bibinfo{pages}{180--196}.
\newblock
\urldef\tempurl%
\url{https://doi.org/10.1287/mnsc.2015.2158}
\showDOI{\tempurl}


\bibitem[\protect\citeauthoryear{Golub and Jackson}{Golub and Jackson}{2009}]%
        {golub2009homophily}
\bibfield{author}{\bibinfo{person}{Benjamin Golub} {and}
  \bibinfo{person}{Matthew~O Jackson}.} \bibinfo{year}{2009}\natexlab{}.
\newblock \showarticletitle{How homophily affects learning and diffusion in
  networks}.
\newblock  (\bibinfo{year}{2009}).
\newblock


\bibitem[\protect\citeauthoryear{Graber and Dunaway}{Graber and
  Dunaway}{2017}]%
        {graber2017mass}
\bibfield{author}{\bibinfo{person}{Doris~A Graber} {and}
  \bibinfo{person}{Johanna Dunaway}.} \bibinfo{year}{2017}\natexlab{}.
\newblock \bibinfo{booktitle}{\emph{Mass media and American politics}}.
\newblock \bibinfo{publisher}{Cq Press}.
\newblock


\bibitem[\protect\citeauthoryear{Jackson and Yariv}{Jackson and Yariv}{2006}]%
        {jackson2006diffusion}
\bibfield{author}{\bibinfo{person}{Matthew~O Jackson} {and}
  \bibinfo{person}{Leeat Yariv}.} \bibinfo{year}{2006}\natexlab{}.
\newblock \showarticletitle{Diffusion on social networks}.
\newblock \bibinfo{journal}{\emph{Economie publique/Public economics}}
  \bibinfo{number}{16} (\bibinfo{year}{2006}).
\newblock


\bibitem[\protect\citeauthoryear{Katz}{Katz}{1957}]%
        {katz1957two}
\bibfield{author}{\bibinfo{person}{Elihu Katz}.}
  \bibinfo{year}{1957}\natexlab{}.
\newblock \showarticletitle{The two-step flow of communication: An up-to-date
  report on an hypothesis}.
\newblock \bibinfo{journal}{\emph{Public opinion quarterly}}
  \bibinfo{volume}{21}, \bibinfo{number}{1} (\bibinfo{year}{1957}),
  \bibinfo{pages}{61--78}.
\newblock


\bibitem[\protect\citeauthoryear{Katz and Lazarsfeld}{Katz and
  Lazarsfeld}{1966}]%
        {katz1966personal}
\bibfield{author}{\bibinfo{person}{Elihu Katz} {and}
  \bibinfo{person}{Paul~Felix Lazarsfeld}.} \bibinfo{year}{1966}\natexlab{}.
\newblock \bibinfo{booktitle}{\emph{Personal Influence, The part played by
  people in the flow of mass communications}}.
\newblock \bibinfo{publisher}{Transaction Publishers}.
\newblock


\bibitem[\protect\citeauthoryear{Kovach and Rosenstiel}{Kovach and
  Rosenstiel}{1999}]%
        {kovach-warp-speed}
\bibfield{author}{\bibinfo{person}{Bill Kovach} {and} \bibinfo{person}{Tom
  Rosenstiel}.} \bibinfo{year}{1999}\natexlab{}.
\newblock \bibinfo{booktitle}{\emph{Warp Speed: {A}merica in the Age of Mixed
  Media}}.
\newblock \bibinfo{publisher}{Century Foundation Press}.
\newblock


\bibitem[\protect\citeauthoryear{Kumar, Mahdian, and McGlohon}{Kumar
  et~al\mbox{.}}{2010}]%
        {kumar-conversations}
\bibfield{author}{\bibinfo{person}{Ravi Kumar}, \bibinfo{person}{Mohammad
  Mahdian}, {and} \bibinfo{person}{Mary McGlohon}.}
  \bibinfo{year}{2010}\natexlab{}.
\newblock \showarticletitle{Dynamics of Conversations}. In
  \bibinfo{booktitle}{\emph{ACM SIGKDD International Conference on Knowledge
  Discovery and Data Mining}}. \bibinfo{pages}{553--562}.
\newblock


\bibitem[\protect\citeauthoryear{Lazarsfeld, Berelson, and Gaudet}{Lazarsfeld
  et~al\mbox{.}}{1944}]%
        {lazarsfeld-peoples-choice}
\bibfield{author}{\bibinfo{person}{Paul~F. Lazarsfeld},
  \bibinfo{person}{Bernard Berelson}, {and} \bibinfo{person}{Hazel Gaudet}.}
  \bibinfo{year}{1944}\natexlab{}.
\newblock \bibinfo{booktitle}{\emph{The People's Choice: How the Voter Makes Up
  His Mind in a Presidential Campaign}}.
\newblock \bibinfo{publisher}{Duell, Sloan, and Pearce}.
\newblock


\bibitem[\protect\citeauthoryear{Molloy and Reed}{Molloy and Reed}{1995}]%
        {molloy1995critical}
\bibfield{author}{\bibinfo{person}{Michael Molloy} {and} \bibinfo{person}{Bruce
  Reed}.} \bibinfo{year}{1995}\natexlab{}.
\newblock \showarticletitle{A critical point for random graphs with a given
  degree sequence}.
\newblock \bibinfo{journal}{\emph{Random structures \& algorithms}}
  \bibinfo{volume}{6}, \bibinfo{number}{2-3} (\bibinfo{year}{1995}),
  \bibinfo{pages}{161--180}.
\newblock


\bibitem[\protect\citeauthoryear{Pariser}{Pariser}{2011}]%
        {pariser-filter-bubble}
\bibfield{author}{\bibinfo{person}{Eli Pariser}.}
  \bibinfo{year}{2011}\natexlab{}.
\newblock \bibinfo{booktitle}{\emph{The Filter Bubble: {What} the {Internet} is
  Hiding from You}}.
\newblock \bibinfo{publisher}{Viking}.
\newblock


\bibitem[\protect\citeauthoryear{Spearman}{Spearman}{1904}]%
        {spearman1904proof}
\bibfield{author}{\bibinfo{person}{Charles Spearman}.}
  \bibinfo{year}{1904}\natexlab{}.
\newblock \showarticletitle{The proof and measurement of association between
  two things}.
\newblock \bibinfo{journal}{\emph{The American journal of psychology}}
  \bibinfo{volume}{15}, \bibinfo{number}{1} (\bibinfo{year}{1904}),
  \bibinfo{pages}{72--101}.
\newblock


\bibitem[\protect\citeauthoryear{Street}{Street}{2010}]%
        {street2010mass}
\bibfield{author}{\bibinfo{person}{John Street}.}
  \bibinfo{year}{2010}\natexlab{}.
\newblock \bibinfo{booktitle}{\emph{Mass media, politics and democracy}}.
\newblock \bibinfo{publisher}{Palgrave Macmillan}.
\newblock


\bibitem[\protect\citeauthoryear{Sunstein}{Sunstein}{2007}]%
        {sunstein-republic-com}
\bibfield{author}{\bibinfo{person}{Cass Sunstein}.}
  \bibinfo{year}{2007}\natexlab{}.
\newblock \bibinfo{booktitle}{\emph{Republic.com}}.
\newblock \bibinfo{publisher}{Princeton University Press}.
\newblock


\bibitem[\protect\citeauthoryear{Sunstein}{Sunstein}{2001}]%
        {sunstein2001echo}
\bibfield{author}{\bibinfo{person}{Cass~R Sunstein}.}
  \bibinfo{year}{2001}\natexlab{}.
\newblock \bibinfo{booktitle}{\emph{Echo chambers: Bush v. Gore, impeachment,
  and beyond}}.
\newblock \bibinfo{publisher}{Princeton University Press Princeton, NJ}.
\newblock


\bibitem[\protect\citeauthoryear{Van~Alstyne and Brynjolfsson}{Van~Alstyne and
  Brynjolfsson}{1996}]%
        {van1996could}
\bibfield{author}{\bibinfo{person}{Marshall Van~Alstyne} {and}
  \bibinfo{person}{Erik Brynjolfsson}.} \bibinfo{year}{1996}\natexlab{}.
\newblock \showarticletitle{Could the Internet balkanize science?}
\newblock \bibinfo{journal}{\emph{Science}} \bibinfo{volume}{274},
  \bibinfo{number}{5292} (\bibinfo{year}{1996}), \bibinfo{pages}{1479}.
\newblock


\bibitem[\protect\citeauthoryear{Watts and Dodds}{Watts and Dodds}{2007}]%
        {watts2007influentials}
\bibfield{author}{\bibinfo{person}{Duncan~J Watts} {and}
  \bibinfo{person}{Peter~Sheridan Dodds}.} \bibinfo{year}{2007}\natexlab{}.
\newblock \showarticletitle{Influentials, networks, and public opinion
  formation}.
\newblock \bibinfo{journal}{\emph{Journal of consumer research}}
  \bibinfo{volume}{34}, \bibinfo{number}{4} (\bibinfo{year}{2007}),
  \bibinfo{pages}{441--458}.
\newblock


\end{thebibliography}
\end{document}